# Z-contrast imaging and *ab initio* study on "*d*" superstructure in sedimentary dolomite

Zhizhang Shen[1], Hiromi Konishi[1], Izabela Szlufarska[2], Philip E. Brown[1], and Huifang Xu[1*]

[1] NASA Astrobiology Institute, Department of Geoscience,

University of Wisconsin - Madison

Madison, Wisconsin 53706

[2] Department of Materials Science and Engineering,

University of Wisconsin-Madison,

Madison, Wisconsin 53706

* Corresponding author: Dr. Huifang Xu

Department of Geoscience,

University of Wisconsin-Madison

1215 West Dayton Street, A352 Weeks Hall

Madison, Wisconsin 53706

Tel: 1-608-265-5887

Fax: 1-608-262-0693

Email: hfxu@geology.wisc.edu

## ABSTRACT

Nano-precipitates with tripled periodicity along the ***c***-axis are observed in a Ca-rich dolomite sample from Proterozoic carbonate rocks with "molar tooth" structure. This observation is consistent with previous description of *d* reflections. High-angle annular dark-field STEM imaging (or Z-contrast imaging) that avoids dynamic diffraction as seen in electron diffraction and high-resolution TEM imaging modes, confirms that *d* reflections correspond to nanoscale precipitates aligned parallel to (*001*) of the host dolomite. The lamellae precipitates have a cation ordering sequence of Ca-Ca-Mg-Ca-Ca-Mg along the ***c*** direction resulting in a chemical composition of $Ca_{0.67}Mg_{0.33}CO_3$. This superstructure is attributed to the extra or *d* reflections, thus is referred to as the *d* superstructure in this study. The structure can be simply described as interstratified calcite/dolomite. The crystal structure of the *d* superstructure calculated from density functional theory (DFT) has a space group of *P31c* and has *a* and *c* unit cell parameters of 4.879 Å and 16.260 Å, respectively, values between those of dolomite and calcite. The detailed structural characteristics and parameters obtained from *ab initio* calculations are also reported in this paper. The method of combining Z-contrast imaging and *ab initio* calculation can be used for solving structures of other nano-precipitates and nano-phases.

## INTRODUCTION

The dolomite($R\bar{3}c$) structure has alternating $Ca^{2+}$ and $Mg^{2+}$ cation layers along the ***c***-axis with the triangular $CO_3^{2-}$ anion layers lying between two cation layers. Since the size of $Mg^{2+}$ ions is smaller than that of $Ca^{2+}$, the $CO_3^{2-}$ layers are closer to $Mg^{2+}$ layers. The lack of *c* glides in the dolomite structure due to the cation ordering causes the occurrence of extra reflections (*b* reflections, $h\bar{h}0l$, $l$= odd) in dolomite diffraction patterns compared to that of calcite (*a* reflections only) (Reeder, 1992). Two additional reflections (*c* and *d* reflections) have been observed in natural sedimentary dolomite. Previously reported *c* reflections were interpreted as being a result of cation ordering within dolomite basal planes (Van Tendeloo et al., 1985). Our recent STEM work confirmed that the "*c*"-reflections could result from multiple diffraction between the host dolomite and twinned Mg-calcite nano-lamellae under TEM imaging and diffraction modes (Shen et al., 2013). The *d* reflection was first observed in Devonian dolomite samples by Wenk and Zenger (1983) and have been also found in ankerite samples (Rekesten, 1990). The *d* reflections occur as satellites around *a* and *b* reflections with diffraction vector ~1/3 (000l)* and are usually streaking along *c** direction (Wenk and Zenger,1983; Wenk and Zhang, 1985; Van Tendeloo et al., 1985).

Dynamical diffraction in transmission electron microscopy (TEM) mode has been a major problem for structure determination as shown in previous work of analyzing "*c*" superstructures (Shen et al., 2013). The scanning transmission electron microscopy (STEM) method uses the high-angle annular dark-field (HAADF) detector to give the most highly localized 1s Bloch state imaging, which eliminates most of the obvious effects of dynamical diffraction (Pennycook, 2002). With the advantage of a spherical aberration corrector, the resolution of HAADF STEM or Z-contrast images is only limited by the size of the 1s Bloch state that is ~0.6-0.8 Å (Pennycook et al., 2000). The intensity of Z-contrast images is dependent on the atomic number of atoms through the $\sim Z^2$ dependence of the Rutherford scattering cross-section, which thus provides chemical information for the material (Kirkland, 1998; Pennycook et al., 2000). A study of microstructures in natural dolomite samples using Z-contrast images may help find answers to previous observations of superstructures and explore new microstructures in dolomite.

In addition to experimental studies in mineralogy, the application of *ab initio* calculations of crystal structure, phase stability, and physical properties of minerals at given pressure and temperature has increased in the past few years (Ogonov et al., 2006; Barnard and Xu, 2008; Chatterjee and Saha-Dasgupta, 2010; Stackhouse et al., 2010). Density functional theory (DFT) uses the functional of electron density to solve the Schrödinger equation for a many electron system to acquire the minimum energy of the system (Scholl and Steckel, 2009). This method can calculate the enthalpy of a system at 0K and the corresponding structure. Since STEM work only provides topological information for the crystal structure, the use of the DFT method is needed in this study to calculate the detailed crystal structure of the *d* phase and confirm our model for the *d* superstructure by comparing the energetic stability of our model to another model proposed in literature.

The "molar tooth" structure refers to vertical and horizontal ribbons and blobs of fine-grained calcite in a dolomite host (Bauerman, 1885). This structure has only been recognized in Mesoproterozoic and Neoproterozoic marine carbonate rocks with a few exceptions (Frank and Lyons, 1998). The origin and the temporally limited occurrence of molar tooth carbonate have bewildered geologists for over a century (James et al., 1998; Pratt, 1998; Meng and Ge, 2002; Marshall and Anglin, 2004; Pollock et al., 2006; Long, 2007; Kuang et al., 2012). However, no previous work on the nano-scale mineralogy of carbonates with the "molar tooth structure" has been conducted. In this present study, STEM imaging and DFT calculations were combined to provide a complete crystallographic description of the *d* superstructure.

## SAMPLES

The "molar tooth" carbonate samples were collected near Hungry Horse Dam, Montana, from outcrops of the Helena Formation of the Mesoproterozoic Belt Group (Frank and Lyons, 1998). The sample HHL-00H was chosen for STEM analysis. In the hand specimen of HHL-00H (Figure 1), the thin sinuous vertical and horizontal ribbons with width of ~0.2-1.5cm intersect each other. The "molar tooth" is composed of clean and fine-grained calcite crystals with similar size (~10 μm). The "molar tooth" host has three major phases: dolomite, calcite, and quartz (Figure 2). K-feldspar, illite, chlorite, rutile, and apatite are also present in the "molar tooth" host rock. The XRD powder analysis of the "molar tooth" host rock shows that the dolomite is cation-ordered with the presence of sharp (105) and (009) peaks. The dolomite has a composition of ~$Ca_{1.02}Mg_{0.98}(CO_3)_2$ based on the $d_{104}$ value (2.890Å) and the relationship between $d_{104}$ values and $MgCO_3$ content in ordered dolomite (Goldsmith and Graf, 1958; Zhang et al., 2010).

## METHODS

Specimens for STEM measurements were prepared by ion milling. Ion milling was performed with a Fischione 1010 ion mill operated at an accelerating voltage of 4 kV and an incident ion-beam angle of 10°, followed by gentler milling at an accelerating voltage of 2.6 kV and an incident angle of 10° in order to reduce surface amorphous material. The ion-milled samples were lightly carbon coated. The microstructures in the dolomite crystals were examined by using a spherical aberration-corrected FEG- STEM (Titan 80-200) operating at 200 kV at the University of Wisconsin-Madison. This instrument can image single atoms with ~ 0.1 nm or better spatial resolution in STEM mode. Probe current was set at 24.5 pA. Camera length for the image acquisition was set at 160 mm. Collection angle of the HAADF detector for acquiring the Z-contrast images ranges from 54 to 270 milliradians (mrads).

The DFT calculations were performed by using the Vienna *ab initio* simulation package (VASP) (Kresse et al., 1996). The general gradient approximation (GGA) with the Perdew, Burke, and Ernzerhof (PBE) parameters was employed (Perdew et al., 1996). The projector-augmented wave (PAW) method with an energy cutoff of 600 eV was used. A conventional hexagonal supercell of calcite derived structures including 30 atoms or 6 chemical formula units was used. We tested k-point convergence and a mesh of 3×3×1 was found to be sufficient for the system. Two starting structures for the *d* phase: one with calcite's unit cell parameters from experiments (Graf, 1969) and another with dolomite's (Beran and Zemann, 1977). All the initial structures were optimized using the static energy minimization scheme, where both the shape and volume of the cell were allowed to relax. The structure with minimum energy calculated from the previous step was further calculated by *ab initio* molecular dynamics simulations at 10K to better explore the local minimum. The powder and electronic diffraction patterns of calculated structures were generated by CrystalDiffract® and SingleCrystal™ respectively.

## HIGH-RESOLUTION (S)TEM OBSERVATIONS

In TEM images, the modulated microstructures with strong strain contrast are prevalent through the "molar tooth" host dolomite. Calcite inclusions that are a common phenomenon in sedimentary dolomites can be easily recognized because of being free of modulations (Figure 3). Diffuse streaks along ***c**** occur in the electron diffraction pattern of the host dolomite (Figure 4). Some maxima of the streaks are about one third of $d^*_{006}$ (Fig. 4, and also see diffraction patterns of Wenk and Zenger, 1983; Wenk and Zhang, 1985). According to the Fast Fourier Transform (FFT) patterns of different areas in the dolomite images, the streaks in the diffraction pattern come from the precipitates with linear features that are parallel to (001*)* (Figure 5). However, the precipitates themselves are arranged in such a way that they are roughly parallel to $(1\bar{1}4)$. The observations above match the features of *d* reflections that were first described by Wenk and Zenger (1983). It was proposed that the streaks parallel to ***c**** could be from stacking disorder (Van Tendeloo et al., 1985).

The dark areas of the precipitates in the bright field (BF) image under STEM mode become bright areas in the HAADF image (Figure 6), which suggests higher Ca contents in the precipitates than the host dolomite. The occurrence of *d* reflections in the FFT pattern of the HAADF image excludes the possibility that they are caused by multiple diffraction. In the HAADF of high magnification or Z-contrast image (Figure 7), the alternating bright and dark layers along ***c*** axis represent the alternating Ca and Mg layers in the normal dolomite structure. However, the *d* dolomite precipitate has the cation sequence of Ca-Ca-Mg-Ca-Ca-Mg- (Figure 7). The FFT pattern of this domain shows that the *d* reflections are attributed to this superstructure. This observation is different from the previous model for the *d* superstructure in which every third Mg layer is replaced by a Ca layer producing a sequence of Ca-Mg-Ca-Ca-Ca-Mg- (Wenk and Zhang, 1985). But both sequences produce the same chemical stoichiometry of $Ca_{0.67}Mg_{0.33}CO_3$. However, the repetition along the ***c***-axis is doubled in the previous model with respect to the observed Ca-Mg ordering in the Z-contrast image (Figure 7).

Mg-bearing calcite precipitates are also observed in the dolomite (Figure 8). Along the $(\bar{1}02)$ trace, the calcite exsolution region has six consecutive Ca layers as opposed to the dolomite host that has alternating Ca and Mg layers. In the line profile of the calcite region, some Ca columns have slightly lower intensities than the pure Ca columns in the dolomite host, which suggests the existence of a small amount of Mg in this calcite precipitate (Figure 8). This calcite exsolution is similar to Mg-calcite precipitates in Ca-rich dolomite (Shen et al., 2013).

## DFT CALCULATIONS

The structural parameters for the optimized *d* superstructure from DFT calculations are listed in Table 1. Tables 2 and 3 compare the calculated lattice parameters of dolomite and calcite structures with the reported experimental data. The calculated equilibrium volumes for both dolomite and calcite are slightly smaller than the reported data, the underestimation of the ***c*** parameter being the major contribution, but are still within the range of previous theoretical calculation data (see Table 2 and 3). The calculated values are for the structures at 0K. The reported experimental values were measured at ambient environment. Temperature could be a factor for the small discrepancy between calculated and measured the unit cell volumes. A small discrepancy between calculated and experimental values of lattice parameters is not uncommon for DFT calculations and may result from the use of an approximate exchange-correlation potential (Hossain et al., 2011). In spite of this discrepancy, the trend found in experimental data is maintained in our DFT calculations. The calculated *d* superstructure has an ***a*** parameter close to that of dolomite but has a ***c*** parameter closer to stoichiometric calcite. This finding is consistent with the observation from the diffraction patterns of the *d* superstructure that the difference between ***a*** parameters of dolomite and the *d* superstructure is smaller than that between ***c*** parameters, even though the DFT calculations are unconstrained bulk structure calculations. This trend is reasonable because superstructure precipitates share the (001) interface with the host dolomite. Smaller differences in the ***a*** dimensions between the host dolomite and d superstructure would cause less strain at the interface.

The C-O bonds in $CO_3^{2-}$ groups are rigid and the C-O distances in both experimental and calculation data are constant no matter what the actual composition of the carbonate mineral is. It is interesting to note that the C-O distances in the *d* superstructure are divided into two categories (Table 1): 1.294Å when a $CO_3^{2-}$ layer is between $Ca^{2+}$ and $Mg^{2+}$ layers along the c axis; and 1.297Å when a $CO_3^{2-}$ layer is in between two $Ca^{2+}$ layers. DFT calculations predict slightly shorter Ca-O and Mg-O bonds in dolomite and calcite than experimental data (Table 2 and 3). The Mg-O distance in the *d* superstructure increases from 2.069Å in calculated dolomite to 2.085 Å. The Ca-O distance in the *d* superstructure differs depending on the oxygen positions; the Ca-O distances (2.374 Å) are larger when the oxygen ions are from $CO_3^{2-}$ group sitting in between $Ca^{2+}$ and $Mg^{2+}$ layers than those (2.355 Å) from $CO_3^{2-}$ group between two $Ca^{2+}$ layers. This is because that the $CO_3^{2-}$ layers are closer to $Mg^{2+}$ layers than $Ca^{2+}$ layers due to the smaller $Mg^{2+}$ radius. The inversion center is missing in the *d* superstructure while the c glide is retained. The existence of (100) and (200) reflections in the diffraction pattern proves that it is not a rhombohedral unit cell, but a primitive hexagonal (Figure 9C). Therefore, the space group is determined to be *P31c* (No. 159). The atom coordinates and symmetry equivalent positions of the *d* superstructure are reported in Table 1.

The superstructure with cation sequence of Ca-Mg-Ca-Ca-Ca-Mg-Ca that was proposed to explain the *d* reflections was referred to as δ structure (Wenk and Zhang, 1985; and Wenk et al., 1991) (see Figure 9A for details). The calculated δ structure has slightly larger unit cell parameters (a = 4.883Å and c = 16.281Å) than the *d* superstructure indicated here. The enthalpy of a unit cell of δ structure is slightly higher than that of the *d* superstructure by 0.54 kJ/mol per $MCO_3$ unit. The previously proposed δ structure is unstable with respect to the current *d* superstructure. The proposed δ structure model was based on the assumption that overlapped diffraction patterns from host dolomite and *d* superstructure were from the precipitates only (Wenk and Zhang, 1985). Careful examination of their diffraction pattern (Fig. 1A of Wenk and Zhang, 1985) and our FFT pattern (Figure 9D), show that the position of 003 is not half way between 002 and 004 of the *d* superstructure, and the position of 009 is not midway between 008 and 0010 of the *d* superstructure. The *d* superstructure does not have reflections with odd *l* due to its *c*-glide. A powder X-ray diffraction pattern with $d_{104}$ value of 2.930Å for the *d* superstructure is also calculated (Fig. 10).

## IMPLICATIONS

The calculated energies of calcite (Ca(CO3)), dolomite ($CaMg(CO_3)_2$), and d superstructure ($Ca_2Mg(CO_3)_3$) are 3620.64, 7132.37 and 10740.22 kJ/mol, respectively. The energy for the *d* superstructure ($Ca_2Mg(CO_3)_3$) is higher (~4.2 kJ/mol per $MCO_3$ unit) than the sum of the calculated energies from dolomite ($CaMg(CO_3)_2$) and calcite ($CaCO_3$) end members. The *d* superstructures serve as a metastable phase with respect to the stoichiometric dolomite and calcite. The *d* superstructures have a maximum of 4 repeats of Ca-Ca-Mg- along the c axis (< 35Å) in our sample and are only stable within dolomite host in the form of nano-precipitates. It is very difficult for extra Ca ions in the dolomite structure to diffuse through carbonate layers and to congregate to form lamellae parallel to the basal plane at low temperature (Wenk et al., 1991; Shen et al., 2013). This explains all the reported occurrences of nano-precipitates of *d*-reflections in natural samples. Wenk et al. (1991) summarized the various proposed superstructures of Ca-Mg carbonates. By using Z-contrast imaging, we can image the cation ordering directly and propose a more accurate structure model. By using the DFT method, we can calculate the detailed structures and explore the energetics of these metastable nano-phases. The methods may be applied to understand many other nano-phases, where there are challenges or artifacts by using other methods.

It was proposed that precursor for the molar tooth dolomite is gelatin-like carbonate mud that is rich in microbial extracellular polymeric substances substance (EPS) (James et al., 1998; Pollock et al., 2006; Long, 2007). Polysaccharides are the dominant components in the EPS. Z-contrast images of the Ca-rich dolomite show two types of Ca-rich precipitates in the host dolomite. Polysaccharides in the carbonate muds can preferentially promote formation of Ca-rich protodolomite (Zhang et al., 2012). Late stage aging during diagenesis and low-grade metamorphism resulted in the observed nano-precipitates of Mg-calcite and the *d* superstructure. Low-temperature non-stoichiometry dolomite with the observed nano-precipitates may be used as a biosignature.

The observed intermediate phases are analogous to those locally ordered domains in mixed-layer clay minerals, such as interstratified chlorite/serpentine, chlorite/biotite, and chlorite/pyrophyllite minerals (Banfield and Bailey, 1996; Xu and Veblen, 1996; Xu et al., 1996; Wang and Xu, 2006). The observed intermediate phase between calcite and dolomite may be described as interstratified calcite/dolomite, instead of δ–dolomite or *d*-dolomite.

## ACKNOWLEDGEMENTS

This work is supported by NASA Astrobiology Institute (N07-5489) and NSF (EAR-095800). Shen also thanks alumni of the Department of Geosciece for supporting his field trips.

Figure 1. (A)The outcrop shows the elastic deformation inside the carbonate layers with “molar tooth” structure. (B) Sinuous dark blue riboons of fine crystallined calcite (“molar tooth”) exist in the dolomite host that is weathered into buff color. (C) Weathered surface of HHL-00H specimen . (D) “Molar tooth”  becomes white color in fresh cleaved surface of HHL-00H specimen.

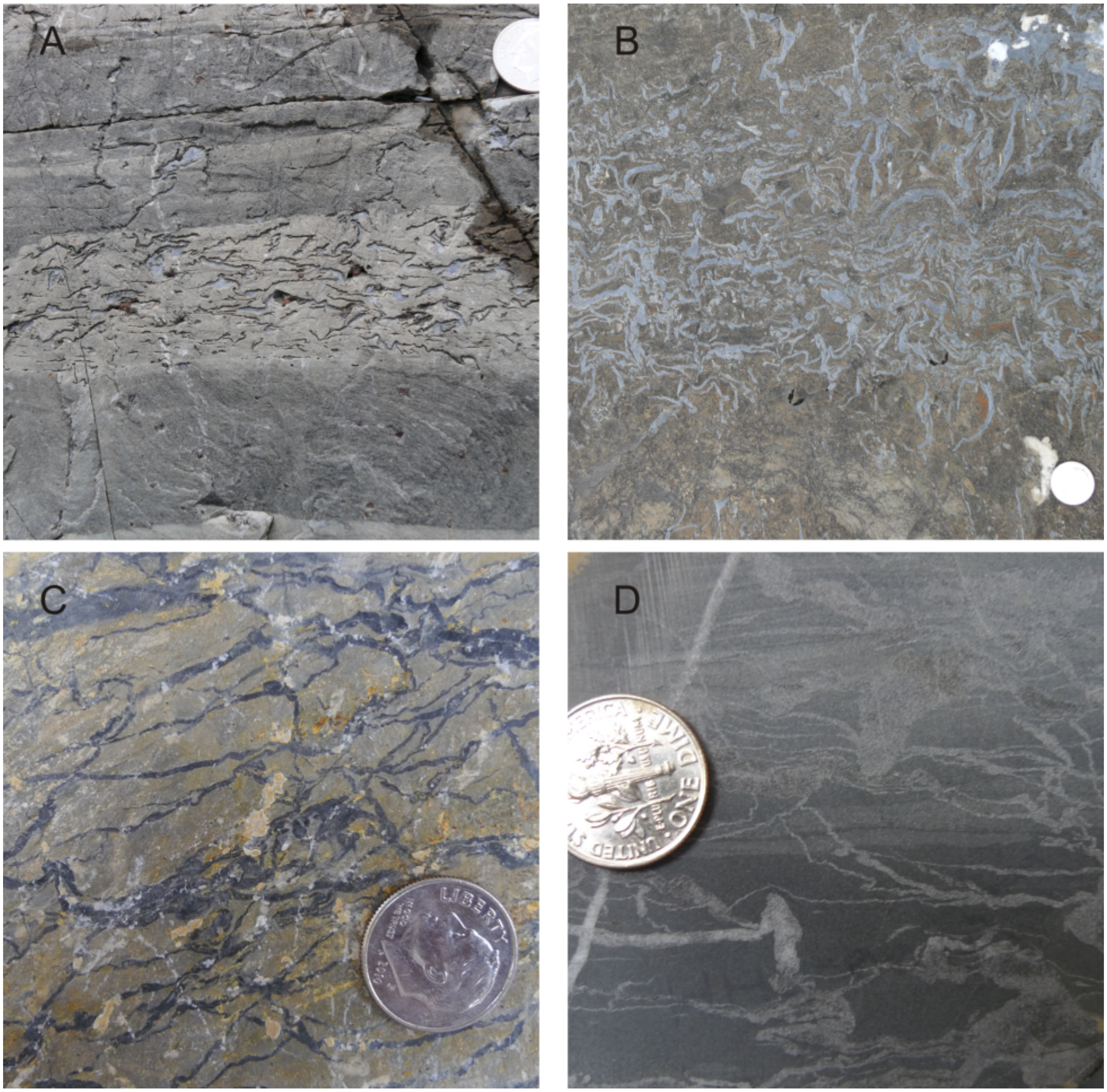

Figure 2. Powder XRD pattern of “molar tooth” host and “molar tooth” in sample HHL-00H.

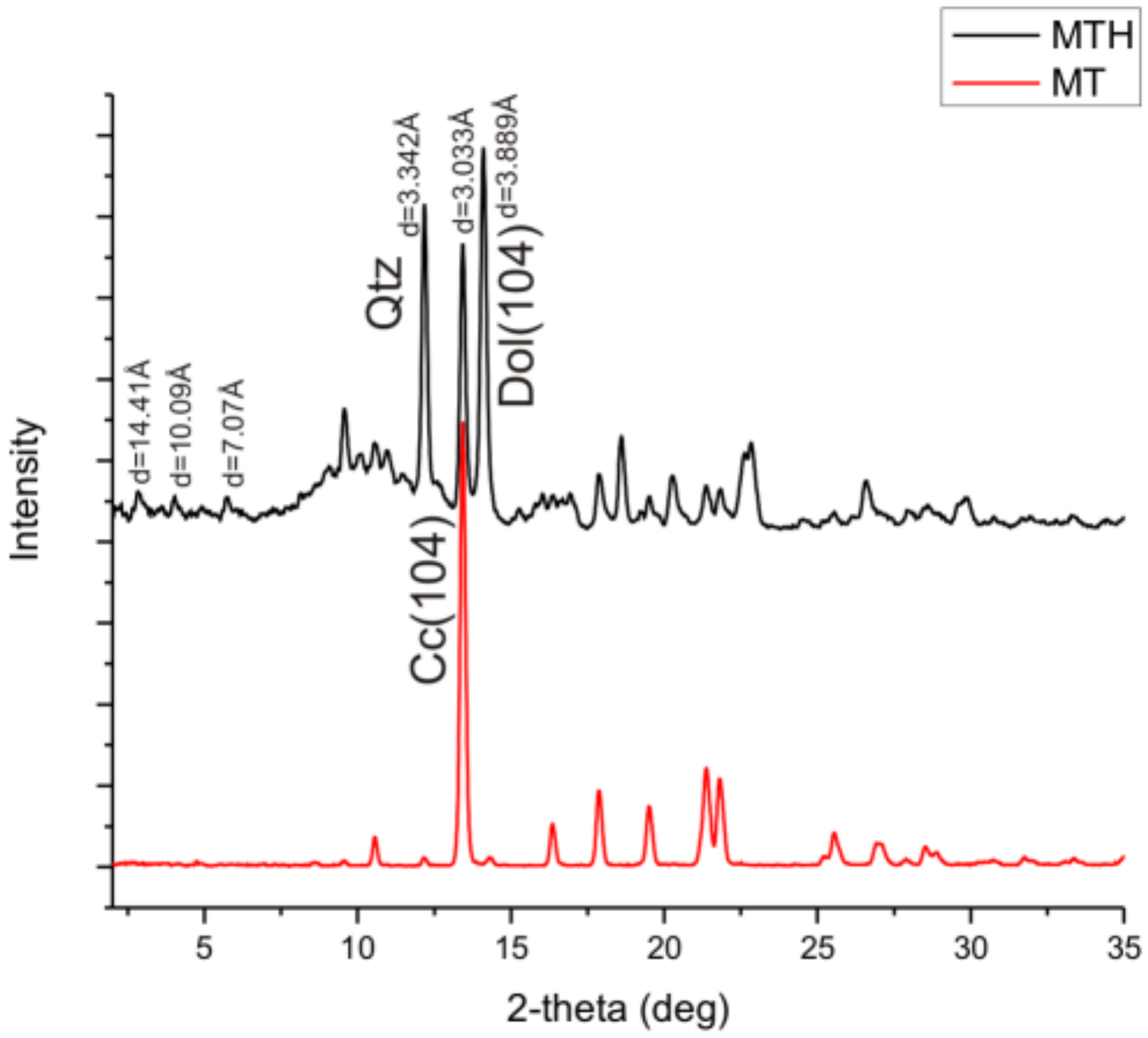

Figure 3. TEM image shows that calcite inclusion that is free of modulations or strain contrast exists in Ca-rich dolomite host that is with a lot of modulations.

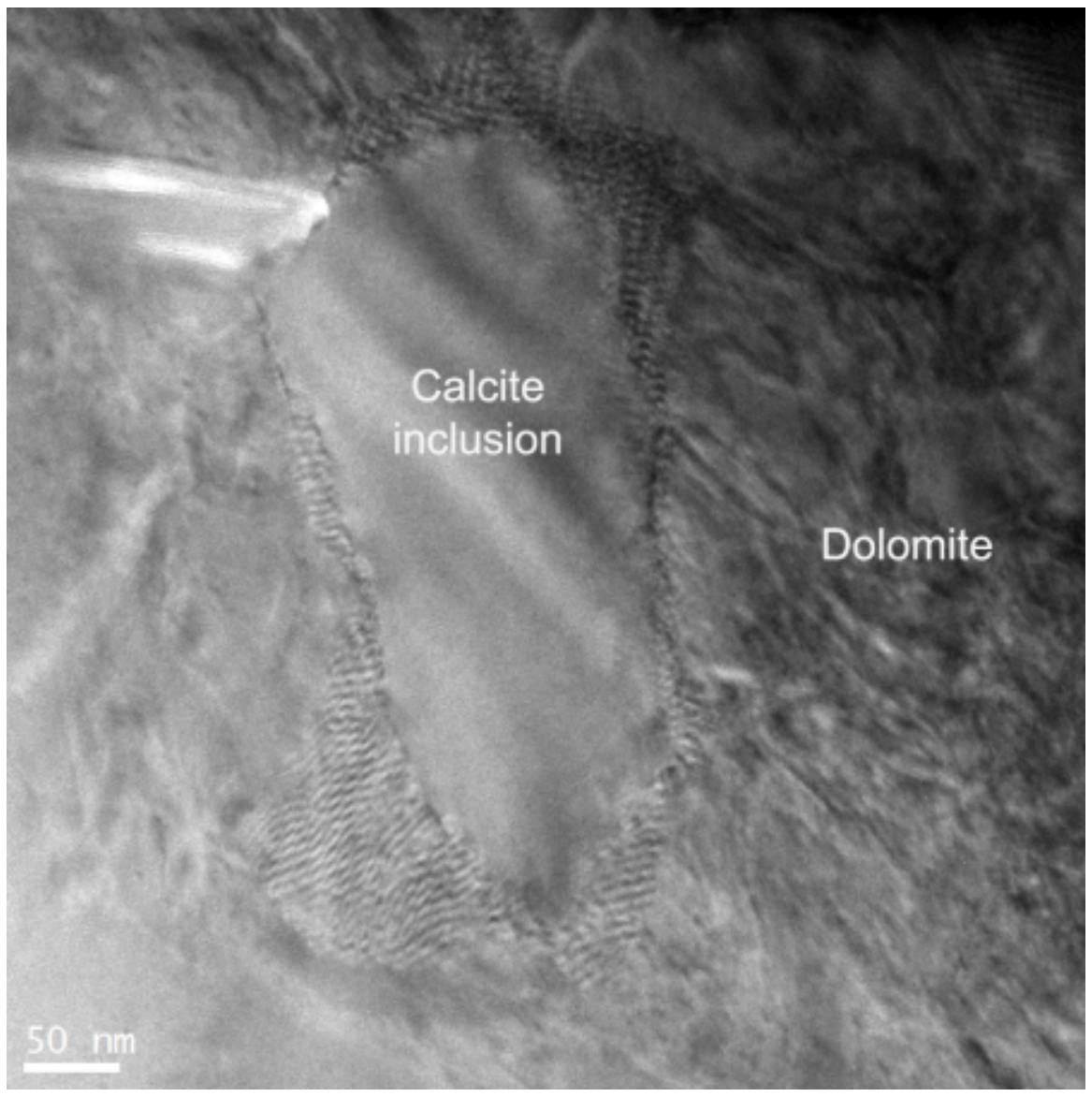

Figure 4. The diffuse streaks along c* occur in the diffraction pattern of Ca-rich dolomite from Helena Formation, Montana. The vector length is roughly around 1/3 of $d^*_{006}$.

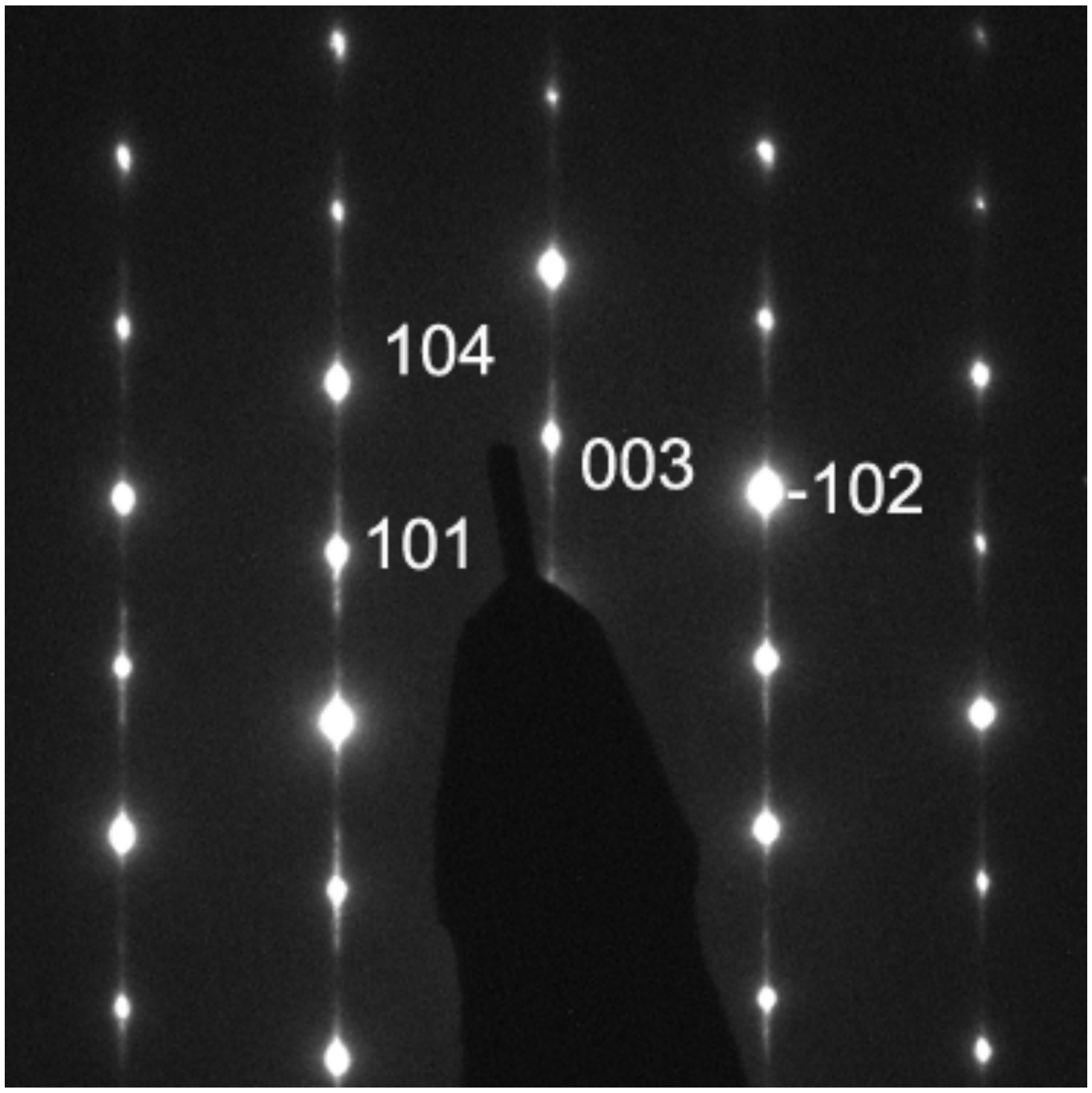

Figure 5. FFT patterns from TEM images show that the “d” reflections are corresponding to domains with //(001) linear feature (compare the FFTs from zone a and zone b). The //(001) modulations are arranged parallel to (104).

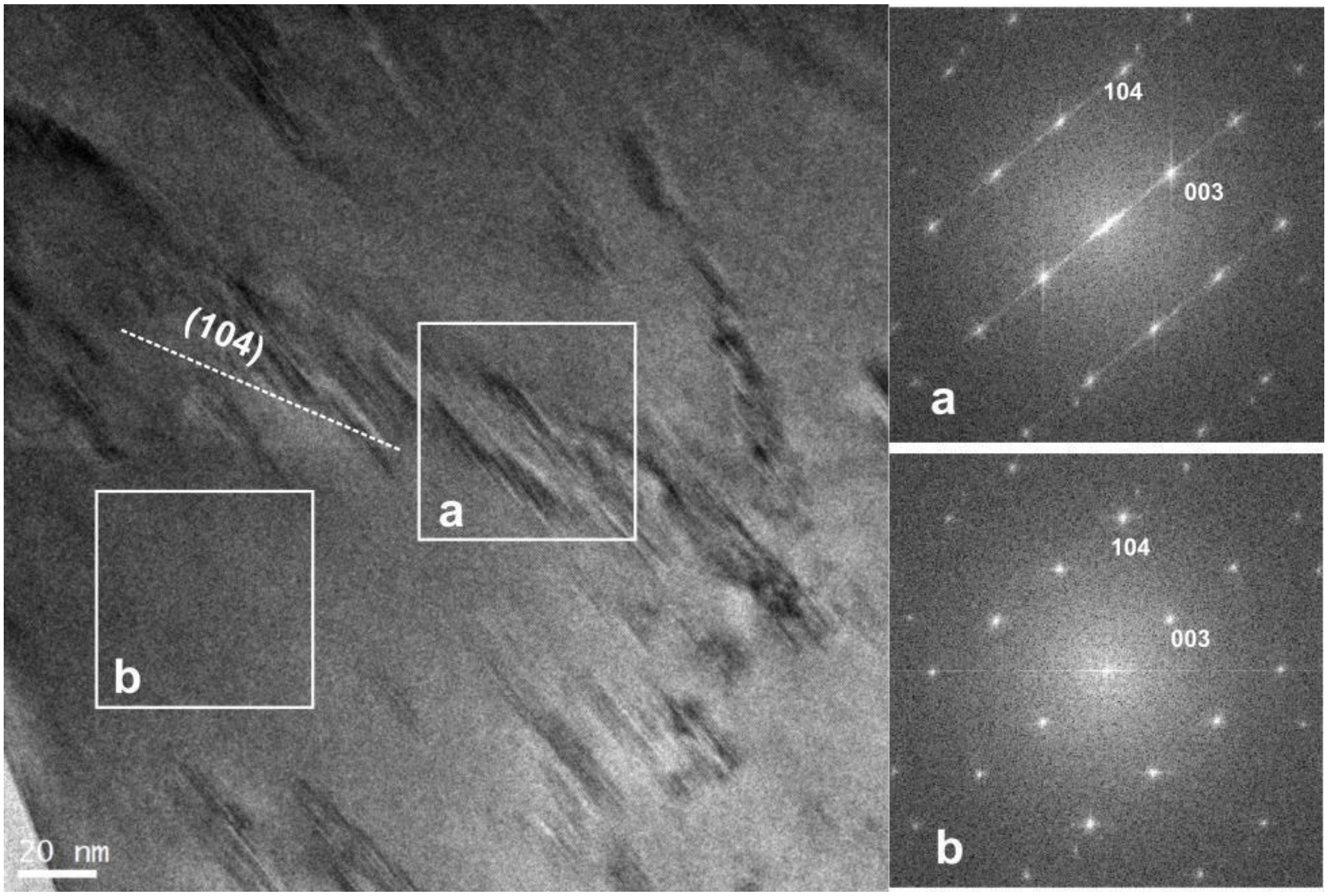

Figure 6. The dark domains of *d* superstructure in the bright-filed image (left) under STEM mode become bright in HAADF image (right), which means a higher Ca contents in the domains of *d* superstructure than the host dolomite.

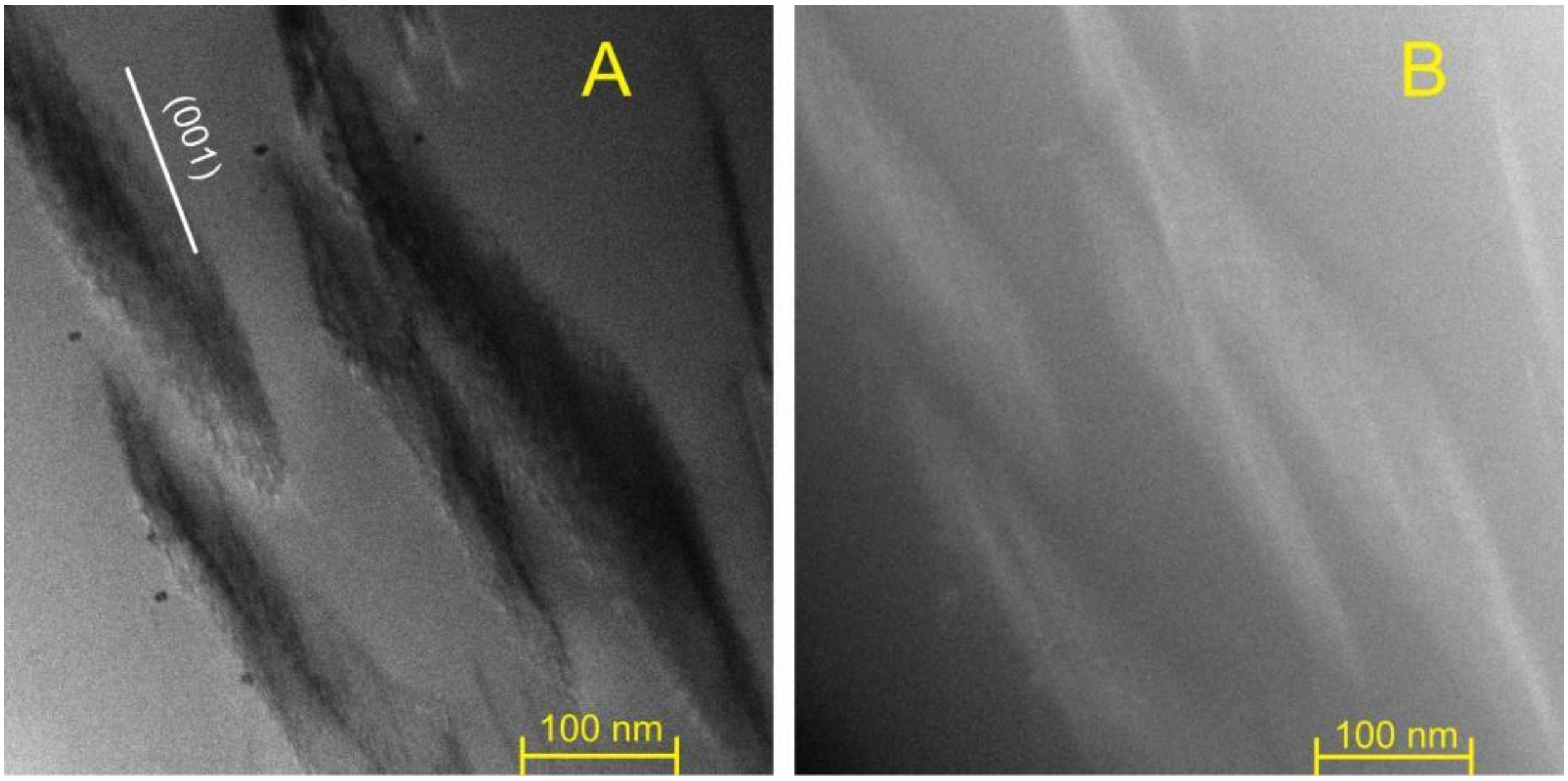

Figure 7. High magnification bright field (A) and dark filed (B) STEM images of *d* domains. Z-contrast (dark filed) images show that *d* domains have a cation sequence (bottom-right) of Ca-Ca-Mg-Ca-Ca-Mg-Ca along *c* axis as opposed to dolomite cation sequence as shown in the middle-right corner. The occurrence of streaking or splitting along *c** or *d* reflections (see the two separate reflections satellite the (003) reflection) in the FFT pattern (bottom-left corner) of the HAADF image excludes the possibility that it is caused by multiple diffraction in TEM mode. In inverse FFT image (C) from FFT pattern of Z-contrast image (B), cation sequence feature is enhanced and double-confirmed in the line profile.

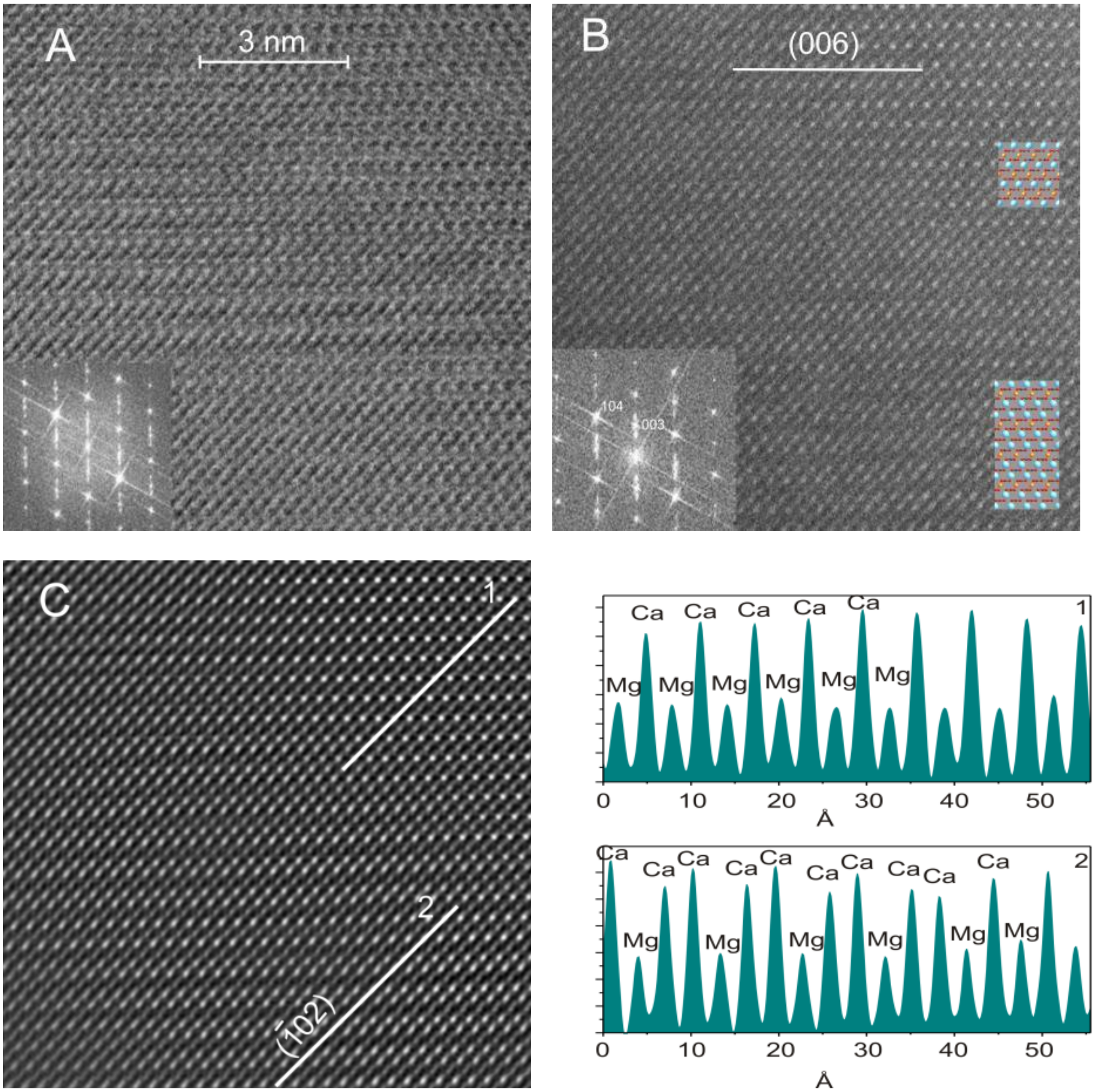

Figure 8. Calcite exsolution lamellae // (001) exist in dolomite host. The line profile 1 has six consecutive Ca columns along $(\bar{1}02)$ trace. The line profile 2 of the dolomite region shows the normal dolomite cation sequence of alternating Ca and Mg columns. The line profile 3 shows one repeat of *d* superstructure. The lines with arrows show the boundaries between calcite and dolomite and between dolomite and *d* superstructure. The atomic models for dolomite/calcite and dolomite/*d* phase interfaces are shown at the bottom. Carbonate group are not shown proportionally in order to highlight the cation sequences.

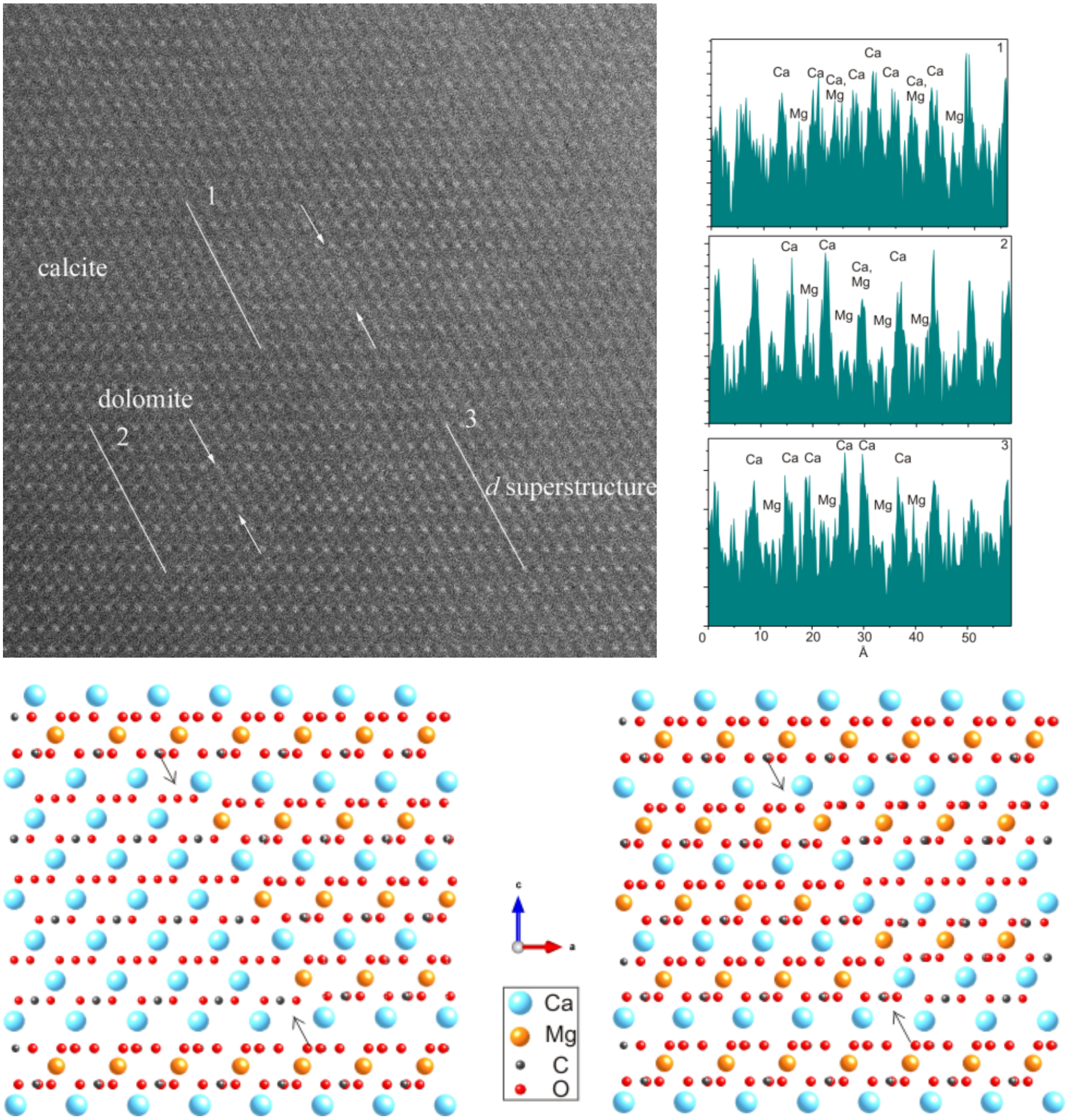

Figure 9. Previously proposed structure model (A), and calculated model for *d* superstructure (B), and corresponding calculated electron diffraction pattern (C), compared to the FFT pattern (D) from the *d* supterstructure, and calculated powder diffraction pattern (E). Very weak 003 and 009 reflections also occur in FFT pattern due to contributions from the dolomite. However, the position of 003 is not at the center of 002 and 004 of the d supterstructure, and the position of 009 is not at the center of 008 and 0010 of the d supterstructure. The d superstructure does not have reflections with odd l due to its *c*-glide.

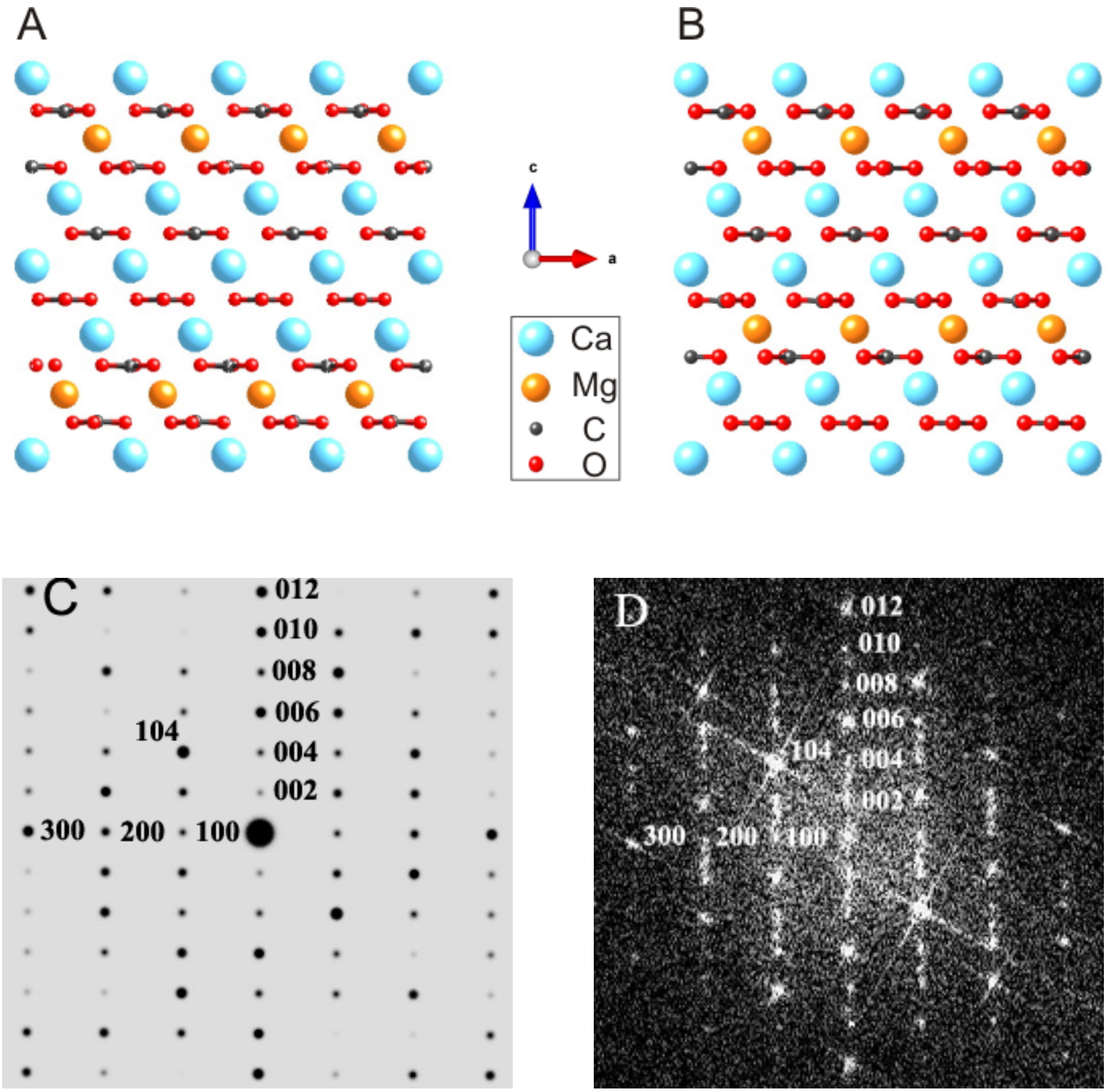

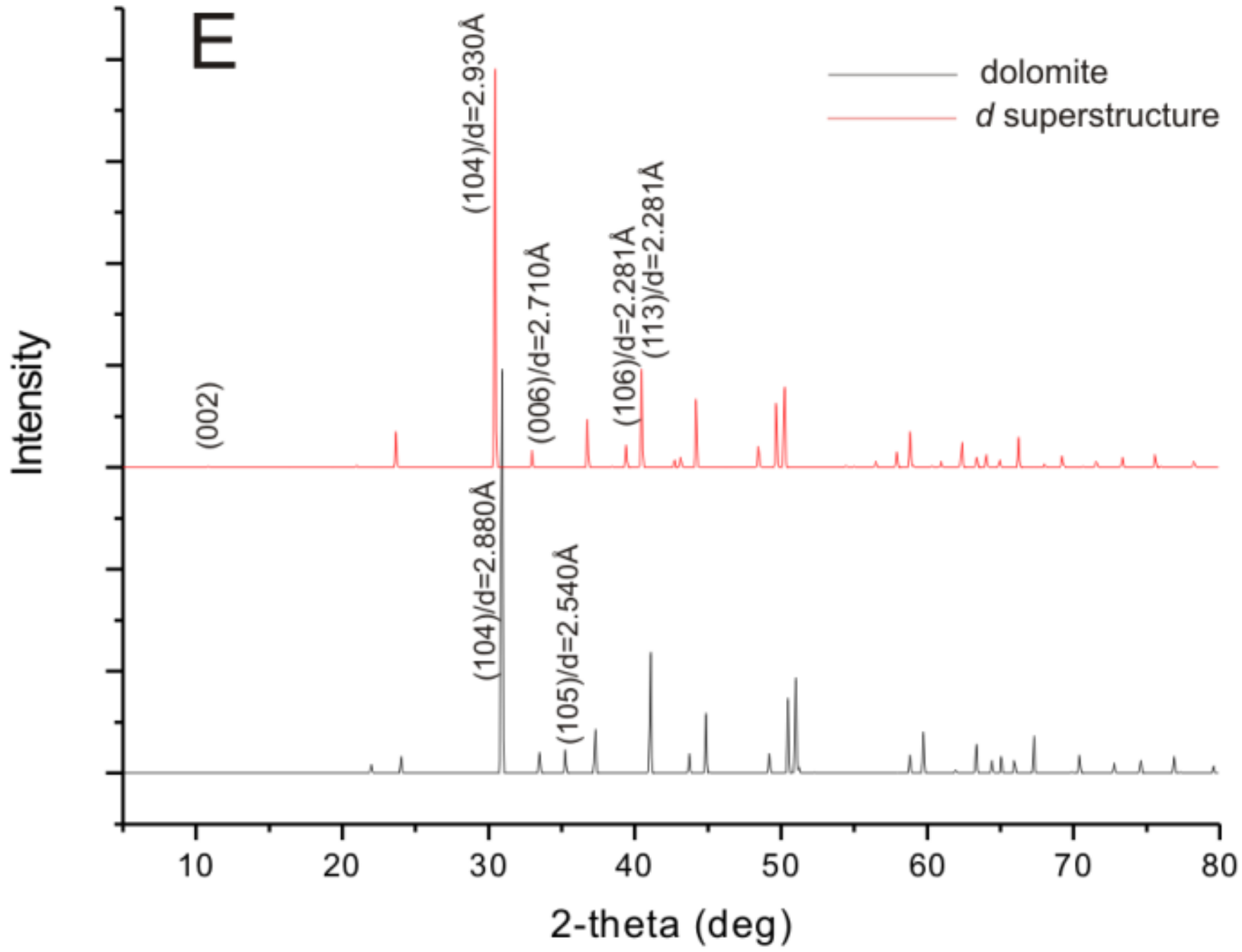

Table1. Lattice parameters for "*d*" superstructure optimized from DFT calculations.

| Lattice parameters | a (Å) | c (Å) | V (Å$^3$) | C-O bond (Å) | Ca-O bond (Å) | Mg-O bond (Å) |
|---|---|---|---|---|---|---|
| This study | 4.879 | 16.260 | 335.226 | 1.294, 1.297 | 2.355, 2.374 | 2.085 |

Table 2. Calculated structural parameters of dolomite as compared to the previous experimental and theoretical data.

| Lattice parameters | This work | Experimental | Theoretical |
|---|---|---|---|
| a (Å) | 4.810 | 4.808[a], 4.812[b] | 4.787[c], 4.877[d], 4.858[e] |
| c (Å) | 15.704 | 16.010[a],16.020[b] | 15.55[c], 16.285[d], 16.109[e] |
| V (Å$^3$) | 314.611 | 320.504[a],321.251[b] | 308.623[c],335.409[d],329.248[e] |
| C-O bond (Å) | 1.294 | 1.233[a], 1.286[b] | 1.286[c], 1.299[d] |
| Ca-O bond (Å) | 2.358 | 2.405[a], 2.382[b] | 2.328[c], 2.401[d] |
| Mg-O bond (Å) | 2.069 | 2.114[a], 2.087[b] | 2.071[c], 2.314[d] |
| O-Mg-O bond angles (°) | 89.03, 90.97, 180 | 89.17, 90.83, 180[a] | 89.335, 90.645, 180[c] 88.546,91.454,180[d] |

[a] Graf ; [b] Beran and Zemann; [c] Hossain et al., LDA functional was used; [d] Hossian et al., GGA functional was used; [e] Bakri and Zaoui, fitted to Birch-Murnaghan equation of state.

Table 3. Calculated structural parameters of calcite as compared to the previous experimental and theoretical data.

| Lattice parameters | This work | Experimental | Theoretical |
|---|---|---|---|
| a (Å) | 5.009 | 4.990[a] | 5.061[b], 4.981[c] |
| c (Å) | 16.614 | 17.062[a] | 17.097[b],15.902[c] |
| V (Å$^3$) | 361.058 | 367.916[a] | 379.279[b], 341.676[c] |
| C-O bond (Å) | 1.295 | 1.286[a] | |
| Ca-O bond (Å) | 2.339 | 2.357[a] | |

[a] Graf; [b] Ayoub, Zaoui, and Berghout; [c] Aydinol et al.

Table 4. Fractional coordinates of atoms in "*d*" superstructure (space group: *P31c* (no. 159)).

| Atom | x | y | z |
|---|---|---|---|
| Ca1 | 0.00000 | 0.00000 | 0.00000 |
| Ca2 | 0.33333 | 0.66667 | 0.18457 |
| Mg | 0.66667 | 0.33333 | 0.34229 |
| C1 | 0.66667 | 0.33333 | 0.09229 |
| C2 | 0.00000 | 0.00000 | 0.26849 |
| C3 | 0.33333 | 0.66667 | 0.41608 |
| O1 | 0.66667 | 0.06803 | 0.09229 |
| O2 | 0.28361 | 0.04018 | 0.26739 |
| O3 | 0.04972 | 0.62648 | 0.41719 |